\documentstyle[preprint,prb,aps]{revtex}

\begin{document}
\draft
\title{Microwave Nonlinearities in High-T$_c$ Superconductors: The Truth Is Out
There}
\author{Steven M. Anlage, Wensheng Hu, C. P. Vlahacos, David Steinhauer, B. J.
Feenstra, Sudeep K. Dutta, Ashfaq Thanawalla, and F. C. Wellstood}
\address{Center for Superconductivity Research, Department of Physics, University of\\
Maryland, College Park, MD 20742-4111 }
\maketitle

\begin{abstract}
This paper discusses some of the major experimental features of microwave
nonlinearity in high temperature superconductors, both intrinsic and
extrinsic. The case is made for solving the problem of extrinsic
nonlinearity through the use of localized measurements of microwave surface
impedance and electromagnetic fields. Along these lines, a brief
introduction is given to our work on scanning near-field microwave
microscopy.
\end{abstract}

\section{Introduction}

There is widespread hope that high-T$_c$ superconducting (HTS) microwave
devices will be the first to fulfill the promise of abundant everyday
applications of superconductors. However, the use of HTS materials in these
applications is predicated on their ability to remain linear under a variety
of operating conditions. This article briefly reviews our experimental
understanding of nonlinearities in HTS materials and microwave devices, and
then proposes a new route to understanding and eventually controlling the
microscopic causes of nonlinearity. The subject of nonlinearity in high-T$_c$
superconductors is truly immense and beyond the scope of this paper to
properly summarize. Instead, we will provide one perspective on this aspect
of high temperature superconductivity and a simple vision for its future.
The reader should be aware that many other perspectives exist on this
subject and the references should be consulted whenever possible for some of
these alternative ideas.

A great deal is known about the properties of HTS materials at microwave
frequencies, and particularly at high power. There exist several excellent
review articles\cite{Newman+Lyons,Sam Rev. Article,HeinReview} and books\cite
{PortisBook,ZYShen book,Lancaster} on the nonlinear properties of high-T$_c$
superconductors, and should be consulted first by newcomers to the field. We
will first review the main experimental techniques used to quantify the
degree of nonlinearity present in superconducting devices and materials. We
then discuss the sources of nonlinearity which have been identified,
concentrating on those sources of nonlinearity which are known to be
extrinsic. We next discuss some empirical correlations between the material
and device properties and the degree of nonlinearity. We then briefly
discuss new techniques designed to characterize inhomogeneities in materials
and seek out the microscopic origins of nonlinearity. Finally we conclude
with our vision for the future evolution of this field.

\section{Experimental Methods}

Nonlinearity in superconducting devices is manifested by the fact that the
device behavior is not invariant with respect to changes in the input power.
Properties which change with power level include the widening of the pass
band frequency range and increase of the insertion loss of a band-pass
filter, or the reduction of the quality factor (Q) and shift in the resonant
frequency of a high-Q superconducting resonator. Other forms of nonlinearity
come from frequency conversion, and include intermodulation of two or more
signals in the pass band of a filter, or harmonic generation in wideband
devices such as delay lines. All of these manifestations of nonlinearity can
be traced to the fact that the electrodynamic properties of a superconductor
are fundamentally nonlinear. One direct measure of nonlinearity is how the
surface impedance depends on rf power. However, more sensitive (and
practical) measurements of nonlinearity exist. These include quantitative
measurements of harmonic generation and intermodulation distortion. All of
these characterization methods are global in the sense that the entire
device is treated as a ``black box''. To make progress in taming
nonlinearities it is essential that we obtain a microscopic picture of the
sources of nonlinearities. In section IV we discuss a new class of
experimental methods which are being developed to uncover the microscopic
origins of material inhomogeneity and nonlinearity.

\subsection{ Power-Dependent Surface Impedance}

Resonant techniques are often employed to measure the surface impedance of
superconductors. The surface resistance can be deduced from measurements of
the Q of the resonator, while the surface reactance is found from
measurements of the frequency shift of the resonator.\cite{Olivier} It is
generally found that the surface resistance and reactance tends to increase
as higher microwave powers are applied. Nonlinearity often shows up first as
changes in the resonator transmission response as a function of frequency.%
\cite{OatesNbN,SridharRes,Snortland,Ma} The surface resistance and reactance
are generally observed to increase quadratically with rf field strength, H$%
_{rf}$, over a range of rf powers.\cite{OatesEarly,OatesLater} Beyond a
critical rf field strength, the increase is faster than quadratic, possibly
indicating heating, or a breakdown associated with the
superconducting-to-normal-state transition. These measurements have the
disadvantage that they are global measurements and make the assumption that
the material is homogeneous. In reality only a few areas of the film may be
dominating the losses.\cite{Hein97} Thus a relatively homogeneous film with
a few defective areas may be judged equivalent to a homogeneously degraded
film.\cite{Reznik}

\subsection{ Harmonic Generation}

A somewhat more sensitive measure of nonlinearity can be obtained through
measurements of harmonic generation in HTS devices. In this method, a signal
at frequency f$_1$ is applied to the device. If the device is time reversal
symmetric (e.g. no trapped flux is present) then the I-V curve must obey the
condition $V(-I)=-V(I)$. If this is the case, then only odd harmonic terms
can arise from nonlinearities in the I-V curve of the device.\cite{ZYShen
book,JiandTinkham} The first (f$_1$) and third (3f$_1$) harmonics exiting
the device are measured as a function of the input power at frequency f$_1$.%
\cite{Golos3f,Wilker3rd,Hampel3f} It is found that with increasing field
strength, harmonic generation and intermodulation, will generally appear
before one observes a change in the surface impedance.\cite{OatesNbN} A
significant disadvantage of this technique is the need to create a broadband
device,\cite{Wilker3rd} or measurement system,\cite{Hampel3f} which can
characterize the sample at both the fundamental and third harmonic
frequencies. In addition, one obtains no detailed spatially resolved measure
of where or how the harmonics are generated, except at the edges.\cite
{Hampel3f}

\subsection{ Intermodulation}

The intermodulation measurement of nonlinearity overcomes the problem of
device bandwidth required for harmonic generation measurements. It also has
the advantage of simulating a common real-world problem associated with
nonlinearity: the mixing of two or more signals to create phantom signals in
the same band. In this technique, two signals at frequencies f$_1$ and f$_2$
are applied to the device, both within the passband. Typically the
difference frequency $\Delta $f = f$_2$-f$_1$ is much less than the device
bandwidth, but large enough to be outside the wings of the phase noise
associated with the primary signals. The third order nonlinearity generates
harmonics at frequencies 2f$_1$-f$_2$ and 2f$_2$-f$_1$(among others), which
are $\Delta $f below f$_1$ and $\Delta $f above f$_2$, respectively. Since
all the signals are within the device bandwidth, no special broadband design
is necessary. As the power in the primary signals is increased, the power
observed in third harmonic signals generally increases three times as fast.%
\cite{ZYShen book} Although this characterization technique is convenient
and directly relevant to such things as filter applications, it suffers once
again from being a global or ``black box'' technique. In other words
conventional intermodulation simply describes the overall nonlinearity and
does not reveal local root causes of the problem.

Another problem with both harmonic generation and intermodulation
measurements is that the harmonic signal often does not scale as the third
power of the input signal. It is often observed that the slope of the third
harmonic response vs. the input power has a slope less than 3, or it
executes an ``S-shaped'' pattern.\cite{Wilker3rd,Diete,ShenASC} A proposal
that the S-shaped third harmonic response is due to a redistribution of
current in the device does not appear to be correct.\cite{Kirsten} Clearly
resolution of these questions will only come from a local investigation of
the sources of the third order nonlinearity.

\section{The Origins of Nonlinearity in HTS Materials}

There are four general categories of effects which cause nonlinearity in HTS
materials. A good visual aid for these effects can be found in the review
article mentioned above.\cite{Sam Rev. Article} Each class is considered
separately below.

\subsection{Intrinsic Nonlinearity of HTS materials}

Superconductors exist because they enjoy a free energy advantage over the
normal state at the same value of temperature, transport current, and
magnetic field. However, this free energy advantage is finite, and can be
reduced to zero by increasing the temperature, the magnetic field, or by
applying a dc transport current. In the last case, a significant amount of
energy can be stored in the kinetic energy of the supercurrent flow, thus
reducing the free energy gain of the superconducting state. The price that
is paid to transport the current is that the superconducting order parameter
is suppressed. This in turn reduces the superfluid density, and hence
increases the magnetic penetration depth. If still more current is carried
by the superconductor, the suppression of superconducting properties is even
more severe, until eventually the free energy advantage of maintaining the
superconducting state is lost entirely. This process is strongly nonlinear
and is intrinsic to the superconducting state.

The practical consequences of these nonlinearities usually occur only under
relatively extreme conditions. Many of the effects are associated with the
non-linear Meissner effect. A nonlinear Meissner effect means that a
superconductor will not generate a magnetization which is strictly
proportional to the applied field (as in the ideal Meissner state).
Furthermore, in a d-wave superconductor, the supercurrent flowing in the
directions of the nodes of the order parameter can more easily gain kinetic
energy sufficient to cause depairing, and reduce the superfluid response.
Because the gap increases for directions on either side of the node, the
amount of depairing will depend on the strength of the applied field. One
consequence of this effect is nonlinearity in the power-dependent inductance
of superconducting transmission lines. For a d-wave superconductor one finds
that the 1 dB compression point for the kinetic inductance of a thin film
(thickness less than the penetration depth) superconducting transmission
line is on the order of 10$^3$ Oe.\cite{Mao} This field scale is very high
compared to H$_{c1}$, but may occur in situations where there is a very
strong field parallel to the superconducting surface, such as in a disk
resonator.

More detailed calculations by Dahm and Scalapino show that intermodulation
and third harmonics are dominated by the nonlinear inductance of a d-wave
superconductor in the intrinsic limit.\cite{Dahm1,Dahm2} The magnitude of
these nonlinearities are very small compared to the more commonly
experienced extrinsic nonlinearities discussed below. One distinctive
prediction of their theory is that in the presence of intrinsic d-wave
nonlinearity, there should be an upturn in the intermodulation output at low
temperatures,\cite{Dahm1,Dahm2} distinctly different from nonlinear s-wave
superconductors, and many other forms of nonlinearity.

Another form of intrinsic nonlinearity may come from spontaneously generated
trapped flux in $\pi $-junctions at high-angle grain boundaries in HTS
materials.\cite{Mannhart} This may give rise to second harmonic generation
in the absence of an externally applied magnetic field.\cite{Humphreys}

\subsection{Extrinsic Nonlinearities in HTS materials}

In principle, all extrinsic sources of nonlinearities in HTS materials can
be controlled by careful fabrication and design. Geometrical sources of
nonlinearity have been largely tamed through the use of novel disk or Corbino%
\cite{JimBooth} resonator geometries. However, HTSC materials display
numerous types of microscopic and mesoscopic defects, and there is still
little detailed understanding of how they contribute to nonlinearity. These
defects include weak links as well as other defects which cause runaway
heating of the material at high powers. The central question which we must
address is this: which defects are responsible for nonlinearity, which can
be safely ignored, and which are actually beneficial? Once one knows which
defect is responsible for the problem its effects on the surface impedance
can be modelled and its presence can be reduced by altering the film growth
process.

\subsubsection{Geometry}

In planar strip conducting structures, the spatial distribution of current
flowing parallel to an edge is influenced by self magnetic fields.\cite
{Sheen} The current density is locally increased at edges\cite{Willemsen}
and, for a given incident power, large screening currents can cause the
field to exceed H$_{c1}$. Thus it is believed that nonlinear response starts
at the edges.\cite{Hampel3f} The well known problem of edge current buildup
and subsequent breakdown of superconductivity in thin films has been
attacked from many angles.\cite{Willemsen,Jin,Lam,Lee} Attempts to improve
the situation by tapering the edges of the film or adding normal material do
not seem to help.\cite{Dahm1}

However, it is found that increasing the width of the conducting strip does
help reduce the amount of nonlinearity at a given power level.\cite
{Wilker3rd} To further increase the power handling capabilities of HTS
devices, the shape of the conductor can be changed from relatively narrow
rectangular strips to patches with dimensions similar in both directions.
With this geometry the current is distributed over a larger conductor
surface and the peak field strength at the edges becomes lower.\cite
{ShenASC,Chaolopka} A further refinement on this idea is the use of circular
disk resonators, which are designed to excite rotationally symmetric TM$%
_{0n0}$ modes. These modes offer a unique property: they carry only radial
components of the current density and thus no current flows parallel to the
edges.\cite{Kaiser} Consequently, the resonators are not degraded by the
``edge effect'' described above. An improvement of the power handling
capability by a factor of about 400, in comparison to modes with edge
current, has been demonstrated.\cite{Chaolopka} Another solution to the edge
current problem is to use cylindrical dielectric resonators,\cite
{ShenSapphire,DieteEUCAS} although they take up more volume than planar
devices.

\subsubsection{Weak Links}

Granular superconductors exhibit enhanced nonlinearity due to ``weak links''
between the superconducting grains.\cite{Sam Rev. Article} A granular HTS
material is often modeled as an array of weakly Josephson-coupled
superconducting grains.\cite{Hylton,Attanasio,McDonald} A Ginzburg-Landau
approach can be employed in which the array is approximated as a continuous
medium characterized by effective parameters (coherence length, penetration
depth, critical fields, etc.). Analogies can then be drawn between granular
HTS and ordinary type-II superconductors. High-frequency losses can also be
treated using a coupled-grain model in which the grains are taken to be
purely inductive and the weak links are modeled using the resistively
shunted junction model.\cite{Herd} For weak fields, the array of
superconducting grains will be screened from the applied field. When the
applied field becomes large enough, flux will begin to penetrate the
intergranular material from the surface.\cite{WosikWL} Thus the nonlinear
behavior seen in granular HTS films is due to the nonlinear inductance and
loss within the weak links. This implies that it is necessary to either
employ single-crystal films or work with small grains which are strongly
coupled to minimize the nonlinearity for high-power device applications.\cite
{Halb1,CCChin,Halb2,Yoshitake}

An interesting extreme case of weak link microwave nonlinearity can be
created by growing an HTS film on a bicrystal substrate to create a single
dominant weak link. Oates and collaborators have studied the nonlinear
surface impedance and intermodulation caused by such an engineered weak link
in a stripline resonator.\cite{OatesJJ1,OatesJJ2} They found that the
resistively shunted junction model is successful at describing the main
features of the data, as expected.

\subsubsection{Thermal Effects}

As materials have improved, a new source of nonlinearity has emerged: local
heating and hot-spot formation. These thermal effects can arise from defects
in the substrate and/or the superconducting material itself, or possibly by
heating in the dielectric.\cite{ShenASC} At high microwave current, the
localized heating generated at defects may drive part of a superconductor to
the normal state.\cite{GoloSatnford,Wosik} If the small normal area blocks
the path of the current, there will be a large local heating effect which
may also drive surrounding material into the normal state.\cite{HotspotSBT}
The surface resistance will increase significantly, and the HTS film will
show nonlinear properties, such as harmonic generation and intermodulation.
One possible source of hot spot formation may be surface roughness, and in
particular, a-axis outgrowths.\cite{Tian} Efforts to deal with these
outgrowths through etching of the surface have been modestly successful.\cite
{Roshko}

It appears that thermal grounding of the substrate is critical for
controlling the effects of both local and global heating.\cite{Hein97} The
differences from sample-to-sample and lab-to-lab in the high power behavior
of HTS samples may be due, at least in part, to differences in thermal
conductivity and the efficiency of heat extraction from the substrates.\cite
{Hein97}

\subsection{Structure/Property Correlations}

In the absence of a microscopic understanding of the origins of
nonlinearity, one can still study empirical correlations between deposition
and treatment conditions of the materials and the resulting properties of
HTS devices made from those materials. As an example, Ma {\it et al.}, found
that films with a small magnetic penetration depth $\lambda $, along with
good orthorhombicity in the films, yields devices which are the most linear.%
\cite{Ma} However, this approach is very time consuming, is difficult to
generalize to other methods of growing films,\cite{Ma} and implicitly
assumes that there are no uncontrolled parameters influencing the results.

Another problem with these empirical correlations is the limited range of
deposition and processing conditions over which they hold. For instance, it
is found that films grown by different techniques (laser ablation,
sputtering, thermal evaporation, etc.) must be separately optimized for some
particular microwave property. In addition, groups which optimize different
macroscopic properties for their films can end up with different microwave
power handling capability. For example, one group may optimize the
intermodulation products of its films because that is of interest for the
end user. It is believed that this in turn optimizes the rf-field dependence
of the magnetic penetration depth $\lambda $. On the other hand, another
group may optimize the high power behavior of the surface resistance R$_s$(H$%
_{rf}$) for their films. These different optimization routes may well lead
to different film microstructure. Clearly a microscopic characterization of
the microwave properties and an understanding of superconducting current
interaction with the microstructure is required.

\subsection{Models of Nonlinearity}

Realistic modelling of nonlinearity can only proceed after the microscopic
mechanisms are identified. In the case of weak links, this process has
already been advanced.\cite{Herd,Portis} The origins of hot spots may
include dust or substrate defects, although no clear candidates are known to
generally occur. Other sources of nonlinearity have not been as carefully
modeled. This includes, among other things, the details of the nucleation
and growth of normal phase at the edge of a current-carrying strip.\cite
{Kozyrev} Another important problem is the creation and motion of vortices
at the edges of current-carrying superconducting strips. The formation of an
rf critical state has been addressed by many authors.\cite
{Sridharcrit,Zhoucrit} Recently the time-dependent Ginzburg-Landau equations%
\cite{Tinkham} have been solved for this case.\cite{Aronson1,Aronson2} Among
other things, it is found that normal regions with multiple flux quanta can
be created and move inside the film even in one half of an rf period.
Further experimental work is required to see which of these models correctly
describes the sources of nonlinearity in HTS materials and devices.

\section{Microscopic Characterization}

In an effort to discover the origins of nonlinearity and inhomogeneity in
HTS materials, microscopic imaging techniques are being developed. These
techniques fall into two classes. The first are local materials property
diagnostic techniques which are designed to characterize the homogeneity and
defect structure of materials. The second class are electromagnetic field
(and microwave current) imaging techniques which are being applied to
operating device structures to understand their high power properties. We
discuss each class in turn.

\subsubsection{Materials Property Imaging}

High quality microwave devices with good power handling properties must
start with high quality homogeneous superconducting materials. It is
important to characterize these materials at the same frequency where they
will eventually be used. In an effort to quantify and resolve the microwave
properties of superconducting materials, far-field,\cite{Martens,Holstein}
and near-field\cite{GolosEarlyRImaging,GusFirstPaper,XiangRxImitation}
microwave microscopy techniques have been developed. In the most common
far-field techniques, mm-waves or light are focused on the sample and the
local response is measured. Confocal resonators have been used to image
losses at frequencies on the order of 100 GHz with a few mm spatial
resolution.\cite{Martens,Holstein} Raman microprobe measurements have been
used to characterize the oxygen loss of patterned high-T$_c$ materials,
however they do not seem to correlate with nonlinearities.\cite{Kirsten}

The near-field techniques utilize an electromagnetic radiation source in
close proximity to the sample of interest. If the sample is less than one
wavelength away, the ordinary concepts of far-field radiation patterns no
longer apply. For instance it has been demonstrated that the spatial
resolution of this sort of imaging technique can be many orders of magnitude
less than the wavelength of the radiation.\cite{Ash+Nichols,AnlageASC} The
spatial resolution is given by the larger of two experimentally controlled
quantities: the probe-sample separation and the characteristic feature size
which determines the field structure in the near field of the radiation
source. Spatial resolutions from 10 $\mu $m\cite{AnlageASC} into the sub-$%
\mu $m range\cite{Xiang0.1micron} have been demonstrated and shown to be
independent of the radiation frequency.\cite{AnlageASC}

Near field microwave measurements have been pursued by many groups over the
years. The earliest work by Soohoo\cite{Sohoo} and Bryant and Gunn\cite
{Bryant+Gunn} used scanned resonators with small apertures to couple to the
sample of interest. The method of Ash and Nichols used an open resonator
formed between a hemispherical and a planar mirror.\cite{Ash+Nichols} They
opened a small hole in the planar mirror allowing an evanescent wave to
leave the cavity (like Soohoo), and scanned a sample beneath it. As this
resembles the configuration now used for near-field scanning optical
microscopy, it is often reported to be the first modern near-field
evanescent electromagnetic wave microscope. Near-field imaging has been
accomplished using evanescent waves from the optical to the microwave range
in coaxial, waveguide, microstrip, and scanning tunneling microscope
geometries.\cite
{Durig,Gutman,Tabib-Azar,XiangFirstPaper,Nunes,Hamers,Seifert,Ichiro,Stranick,Keilman,Fee+Chu}

Over the past two years, our group has developed a novel form of near-field
scanning microwave microscopy.\cite{GusFirstPaper} As shown in Fig. \ref
{Schematic}, our microscope consists of a resonant coaxial cable which is
weakly coupled to a microwave generator on one end through a decoupling
capacitor C$_d$, and coupled to a sample through an open-ended coaxial probe
on the other end. As the sample is scanned beneath the probe, the
probe-sample separation will vary (depending upon the topography of the
sample), causing the capacitive coupling to the sample, C$_x$, to vary. This
will result in a change of the resonant frequency of the coaxial cable
resonator because C$_x$ affects the electrical length of the resonator.
Also, as the local sheet resistance (R$_x$) of the sample varies, so will
the quality factor, Q, of the resonant cable because the sample (together
with C$_x$) acts as a terminating impedance for the resonator. A low
frequency feedback circuit is used to force the microwave generator to
follow a single resonant frequency of the cable, and a second circuit is
used to measure the Q of the resonant microscope, both in real time.\cite
{Dave1,Dave2} As the sample is scanned below the open-ended coaxial probe,
the frequency shift and Q signals are collected simultaneously, and
corresponding two-dimensional images of the sample topography and sheet
resistance can be generated. The circuit runs fast enough to accurately
record at scan speeds of up to 25 mm/sec.

We have demonstrated that our scanning near-field microwave microscope can
be used to obtain quantitative topographic images with 55 nm vertical
sensitivity,\cite{GusTopo} and quantitative surface sheet resistance images
of metallic thin films.\cite{Dave2} Figure \ref{YBCOWafer} shows a microwave
frequency sheet resistance image of a YBa$_2$Cu$_3$O$_7$ thin film deposited
on a 2-inch diameter sapphire wafer.\cite{ColorPics} The film shows a sheet
resistance which varies by a factor of 3 over its surface. The image has a
spatial resolution of approximately 500 $\mu $m, and was obtained at room
temperature in only 10 minutes. Such a simple technique could provide a
rapid screening method for wafer-scale homogeneity.

A microwave microscope can probe the material on many different length
scales, allowing one to address a host of issues from large scale
homogeneity to the effect of sample microstructure on the microwave
properties. Higher spatial resolution images have been achieved using
coaxial probes with inner conductor diameters of 200 $\mu $m, 100 $\mu $m,
and 10 $\mu $m. We have also demonstrated the ability to image with an
STM-tip center conductor coaxial cable probe. With this probe we have
demonstrated a spatial resolution on the order of 1 $\mu $m or better while
in contact with the surface. Our microscope also has a broad frequency
coverage (100 MHz to 50 GHz) allowing us to explore the frequency dependence
of microwave properties and, in principle, to perform conductivity depth
profiling experiments.

\subsubsection{Electromagnetic Field and Current Imaging}

Near-field microwave microscopes can also be used to image electromagnetic
fields. Several groups have demonstrated electric field\cite
{Rebeiz,vdW,Ashfaq} and magnetic field\cite{vdWMagnetic} imaging above
operating microwave devices using near-field techniques. In addition several
groups have developed (far field) scanning laser\cite{Kaiser,Newman} and
electron\cite{LTSEM} microscopes which can image microwave currents on the
few $\mu $m length scale. Images from such systems can form the basis for
investigating the interaction of the microwave currents with the
microstructure of the materials making up the device.

Our microscope can also be used to image electric and magnetic fields in the
vicinity of operating microwave devices. Figure \ref{PickUp} illustrates how
we use the open-ended coaxial probe to pick-up electric fields above an
operating device. The microscope detects the normal component of electric
field, i.e. the electric field integrated over the exposed area of the
center conductor of the probe. Thus the spatial resolution is again limited
by the same geometrical parameters which are present in materials diagnostic
mode. The picked up signals are then stored in the resonant system,
rectified by a diode, and recorded on a computer.\cite
{AnlageASC,Ashfaq,Sudeep}

We have imaged a variety of devices using this system. For example, Figure 
\ref{Microstrip}(a) shows a short microstrip transmission line
lithographically defined on a two-sided copper printed circuit board which
terminates at an open circuit on the right hand side. The strip is
approximately 1 mm wide, and the dielectric is 0.5 mm thick. Figure \ref
{Microstrip}(b) shows three active mode images of the microstrip taken at
8.03, 9.66 and 11.29 GHz with a signal applied through the coaxial cable on
the left. The image is taken over the right two-thirds of the microstrip
shown in Fig. \ref{Microstrip}(a). In each case there is a maximum signal on
the right hand side at the position of the open termination demonstrating
that, as expected, the microscope is sensitive to the absolute magnitude of
the electric field rather than current. A clear standing wave pattern is
seen in each image, with a wavelength which decreases linearly with
increasing frequency. Analysis of the three images gives effective
dielectric constants for the printed circuit board microstrip in the range
of 3.32 to 3.46. We have also developed a technique to determine the
approximate value of the perturbed electric field measured by this
microscope.\cite{Ashfaq,Sudeep}

The microscope is sensitive to the component of electric field normal to the
microstrip circuit plane. The image (Fig. 4b) demonstrates clearly that
there are standing waves on the transmission line, and will be of great
assistance in the design of new planar microwave devices (both normal and
superconducting). A cryogenic version of this microscope has been used to
image the electric fields above a superconducting Tl$_2$Ba$_2$CaCu$_2$O$_8$
microstrip resonator operating at 8 GHz and 77 K.\cite{Ashfaq}

There are a number of technical issues which need to be addressed before we
can exploit these microscopes to their fullest potential. The first concerns
the perturbation on the device caused by the presence of the measurement
probe. This perturbation may be so great as to change the current
distributions and obscure the actual behavior. A second issue concerns the
challenge of simultaneously obtaining the required spatial resolution,
topographic information, low temperature operation, and quantitative imaging
capability on an operating microwave device. Both of these challenges call
for continued research by the microwave microscopy community.

\section{Conclusions}

The elimination or control of nonlinearities in HTS materials will require a
microscopic understanding of the defects and geometrical features which
produce extrinsic forms of nonlinearity. We believe that the best route to
this goal is through a quantitative microscopic imaging of materials
properties, electromagnetic fields, and currents, at microwave frequencies
and cryogenic temperatures. Near-field scanning microwave microscopy is
uniquely suited to achieve this goal. We look forward to continued progress
in achieving higher spatial resolution with quantitative cryogenic imaging
capabilities.

The research opportunities for microwave microscopes are virtually
unlimited. We hope that it will soon be possible to combine electromagnetic
field imaging with simultaneously acquired topographic information\cite{vdW}
in the superconducting state. This will allow us to see for the first time
the interactions between microwave currents and the rich and widely varying
microstructure of the HTS materials. This in turn may allow us to understand
which defects are specifically responsible for particular nonlinearity
features. It may then be possible to eliminate or mitigate the effects of
those particular defects to improve the power handling capabilities of HTS
devices.

The use of ideal model systems to understand and model the microscopic
origins of nonlinearity is particularly important. There has been tremendous
progress made in the understanding of Josephson vortex motion at high
frequencies through the use of single grain boundary weak link stripline
resonators. Similar experiments can be done with other idealized sources,
such as thermal defects, tailored edge geometries, and multiple weak link
structures. These experiments need to be accompanied by detailed models
which explain the basic physics of the nonlinearity process. Armed with this
information, it is then possible, in principle, to create films which do not
suffer from those forms of extrinsic nonlinearity and to achieve the goal of
HTS microwave applications free of extrinsic limitations.

Acknowledgments:

We would like to acknowledge many important conversations on this subject
with J\"{u}rgen Halbritter and R. B. Hammond. This work was sponsored by the
National Science Foundation through grant \# ECS-96-32811, and the
NSF/Maryland MRSEC grant \# DMR-96-32521. Additional support has come from
the Maryland Center for Superconductivity Research.

\begin{figure}[tbp]
\caption{Schematic of our scanning near-field microwave microscope. The
microwave source is frequency locked to one of the resonant modes of the
coaxial transmission line resonator. The sample perturbs one end of the
resonator, and the resulting change in resonant frequency and quality factor
are recorded as a function of the position of the probe over the sample..}
\label{Schematic}
\end{figure}

\begin{figure}[tbp]
\caption{Calibrated sheet resistance image of an YBa$_2$Cu$_3$O$_7$ thin
film deposited on a 2-inch diameter sapphire wafer. The image was acquired
at 7.5 GHz with a 480 $\mu $m diameter probe at a height of 50 $\mu $m at
room temperature.}
\label{YBCOWafer}
\end{figure}

\begin{figure}[tbp]
\caption{Schematic illustration of scanning near-field microwave microscope
operated as an electric field microscope over an active device. The
open-ended coaxial probe is sensitive of the electric field normal to the
surface of the inner conductor.}
\label{PickUp}
\end{figure}

\begin{figure}[tbp]
\caption{a) Optical photograph of a copper microstrip circuit showing
coax-to-microstrip transition on left, and open circuit termination on
right. The printed circuit board is 39 mm long and 33 mm wide, and the line
is approximately 1 mm wide. b) The lower part of the figure shows active
images of the copper microstrip with standing waves present. Operating
frequencies are (from top to bottom) 8.03 GHz, 9.66 GHz, and 11.29 GHz. The
horizontal scale is 32.5 mm and the vertical scale on each panel is 3 mm.}
\label{Microstrip}
\end{figure}


\begin{references}
\bibitem{Newman+Lyons}  N. Newman and W. G. Lyons, J.\ Supercond. {\bf 6},
119 (1993).

\bibitem{Sam Rev. Article}  T. B. Samoilova, Supercon. Sci. Technol. {\bf 8}%
, 259 (1995).

\bibitem{HeinReview}  M. A. Hein, in {\it Studies of High-Temperature
Superconductors}, vol. 18, A. Narlikar, Ed. (Nova Sciences, New York, 1996),
p. 141.

\bibitem{PortisBook}  Alan M. Portis, {\it Electrodynamics of
High-Temperature Superconductors}, (World Scientific, Singapore, 1993).

\bibitem{ZYShen book}  Zhi-Yuan Shen, {\it High Temperature Superconducting
Microwave Circuits}, (Artech House, Boston, 1994).

\bibitem{Lancaster}  M. J. Lancaster, {\it Passive Microwave Device
Applications of High Temperature Superconductors}, (Cambridge University
Press, 1997).

\bibitem{Olivier}  O. Klein, S. Donovan, M.\ Dressel, and G. Gr\"{u}ner,
Int. J. Infrared and Millimeter Waves {\bf 14}, 2423 (1993); S. Donovan, O.
Klein, M.\ Dressel, K. Holczer, and G. Gr\"{u}ner, Int. J. Infrared and
Millimeter Waves {\bf 14}, 2459 (1993); M.\ Dressel, O. Klein, S. Donovan,
and G. Gr\"{u}ner, Int. J. Infrared and Millimeter Waves {\bf 14}, 2489
(1993).

\bibitem{OatesNbN}  C. C. Chin, D. E. Oates, G. Dresselhaus, and M. S.
Dresselhaus, Phys. Rev. B {\bf 45}, 4788 (1992).

\bibitem{SridharRes}  B. A. Willemsen, J. S. Derov, J. H. Silva, and S.
Sridhar, IEEE Trans. Appl. Supercon. {\bf 5}, 1753 (1995); T. Jacobs, B. A.
Willemsen, and S. Sridhar, Rev.\ Sci. Instrum. {\bf 67}, 3757 (1996).

\bibitem{Snortland}  H. J. Snortland, Ph.D. Thesis, Stanford University,
1997.

\bibitem{Ma}  Z. Ma, {\it et al.}, IEEE Trans. Appl. Supercon. {\bf 7}, 1911
(1997).

\bibitem{OatesEarly}  D. E. Oates, A. C. Anderson, and P. M. Mankiewich, J.
Supercond. {\bf 3}, 251 (1990).

\bibitem{OatesLater}  D. E. Oates, {\it et al.}, J. Supercond. {\bf 5}, 363
(1992).

\bibitem{Hein97}  M.\ A. Hein, {\it et al.}, J. Supercond. {\bf 10}, 109
(1997).

\bibitem{Reznik}  A. N. Reznik, IEEE Trans.\ Appl. Supercond. {\bf 7}, 1474
(1997).

\bibitem{JiandTinkham}  L. Ji, R. H. Sohn, G. C.\ Spalding, C. J. Lobb, and
M. Tinkham, Phys. Rev. B {\bf 40}, 10 936 (1989).

\bibitem{Golos3f}  M. Golosovsky, D. Davidov, E. Farber, T. Tsach, and M.
Schieber, Phys. Rev. B {\bf 43}, 10 390 (1991).

\bibitem{Wilker3rd}  C. Wilker, {\it et al}., IEEE Trans. Appl. Supercond. 
{\bf 5}, 1665 (1995).

\bibitem{Hampel3f}  G. Hampel, {\it et al}., Appl. Phys. Lett. {\bf 71},
3904 (1997).

\bibitem{Diete}  W. Diete, {\it et al}., IEEE\ Trans.\ Appl. Supercon. {\bf 7%
}, 1236 (1997).

\bibitem{ShenASC}  Z. -Y. Shen, {\it et al.}, IEEE\ Trans. Appl. Supercon. 
{\bf 7}, 2446 (1997).

\bibitem{Kirsten}  K. E. Myers, {\it et al}., IEEE Trans. Appl. Supercond. 
{\bf 7}, 2126 (1997).

\bibitem{Mao}  J. Mao, S. M. Anlage, J. L. Peng and R. L. Greene, IEEE
Trans. Appl. Supercond. {\bf 5}, 1997 (1995).

\bibitem{Dahm1}  T. Dahm and D. J. Scalapino, J. Appl. Phys. {\bf 81}, 2002
(1997).

\bibitem{Dahm2}  T. Dahm and D. J. Scalapino, J. Appl. Phys. {\bf 82}, 464
(1997).

\bibitem{Mannhart}  J. Mannhart, {\it et al}., Phys. Rev. Lett. {\bf 77},
2782 (1996).

\bibitem{Humphreys}  Richard Humphreys, private communication, 1998.

\bibitem{JimBooth}  J. C. Booth, D. H. Wu, and Steven M.\ Anlage, Rev. Sci.
Instrum. {\bf 65}, 2082 (1994).

\bibitem{Sheen}  D. M. Sheen, {\it et al}., IEEE Trans. Appl. Supercond. 
{\bf 1}, 108 (1991).

\bibitem{Willemsen}  B. A. Willemsen, T. Dahm and D. J. Scalapino, Appl.
Phys. Lett. {\bf 71}, 3898 (1997).

\bibitem{Jin}  B. B. Jin, {\it et al}., Supercond. Sci. Technol. {\bf 8},
564 (1995).

\bibitem{Lam}  C. W. Lam, D. M. Sheen, S. M. Ali, and D. E. Oates, IEEE
Trans. Appl. Supercond. {\bf 2}, 58 (1992).

\bibitem{Lee}  L. H. Lee, S. M. Ali, and W. G. Lyons, IEEE Trans. Appl.
Supercond. {\bf 2}, 49 (1992).

\bibitem{Chaolopka}  H. Chaloupka, M. Jeck, B. Gurzinski and S. Kolesov,
Electronics Letters, {\bf 32}, 1735, (1996).

\bibitem{Kaiser}  T. Kaiser, M. A. Hein, G. M\"{u}ller, and M. Perpeet,
submitted to Applied Physics Letters, 1998.

\bibitem{ShenSapphire}  Z. -Y. Shen, {\it et al.}, IEEE\ Trans. Microwave
Theory Tech. {\bf 40}, 2424 (1992).

\bibitem{DieteEUCAS}  W. Diete, {\it et al.}, 1995 EUCAS Conference
Proceedings.

\bibitem{Hylton}  T. L. Hylton, {\it et al}., Appl. Phys. Lett. {\bf 53},
1343 (1988).

\bibitem{Attanasio}  C. Attanasio, L. Maritato, and R. Vaglio, Phys. Rev. B 
{\bf 43}, 6128 (1991).

\bibitem{McDonald}  J. McDonald and John R. Clem, Phys. Rev. B, {\bf 56},
14723 (1997).

\bibitem{Herd}  J. S. Herd, D. E. Oates, and J. Halbritter, IEEE Trans.
Appl. Supercon. {\bf 7}, 1299 (1997).

\bibitem{WosikWL}  J. Wosik, {\it et al}., Phys. Rev. B {\bf 51}, 16 289
(1995).

\bibitem{Halb1}  J. Halbritter, J. Appl. Phys. {\bf 68}, 6315 (1990).

\bibitem{CCChin}  C. C. Chin, D. E. Oates, G. Dresselhaus, and M. S.
Dresselhaus, Phys. Rev. B {\bf 45}, 4788 (1992).

\bibitem{Halb2}  J. Halbritter, J. Supercon. {\bf 8}, 691 (1995).

\bibitem{Yoshitake}  T.Yoshitake, H. Tsuge, and T. Inui, IEEE Trans.\ Appl.
Supercond. {\bf 5}, 2571 (1995).

\bibitem{OatesJJ1}  D. E. Oates, {\it et al}., Appl. Phys. Lett. {\bf 68},
705 (1996).

\bibitem{OatesJJ2}  T. C. L. G. Sollner, J. P. Sage, and D. E. Oates, Appl.
Phys. Lett. {\bf 68}, 1003 (1996).

\bibitem{GoloSatnford}  M. A. Golosovsky, H. J. Snortland and M. R. Beasley,
Phys. Rev. B {\bf 51}, 6462 (1995).

\bibitem{Wosik}  J. Wosik, {\it et al.}, J. Supercond. {\bf 10}, 97 (1997).

\bibitem{HotspotSBT}  W. J. Skocpol, M. R. Beasley, and M. Tinkham, J. Appl.
Phys. {\bf 45}, 4054 (1974).

\bibitem{Tian}  Y. J. Tian, {\it et al}., Appl. Phys. Lett. {\bf 65}, 2356
(1994).

\bibitem{Roshko}  A. Roshko, {\it et al}., IEEE Trans.\ Appl. Supercond. 
{\bf 5}, 1733 (1995).

\bibitem{Portis}  A. M. Portis, Appl. Phys. Lett. {\bf 58,} 307 (1991).

\bibitem{Kozyrev}  A. B. Kozyrev, {\it et al.}, Supercond. Sci. Technol. 
{\bf 7}, 777 (1994).

\bibitem{Sridharcrit}  S. Sridhar, Appl. Phys. Lett. {\bf 65}, 1054 (1994).

\bibitem{Zhoucrit}  Shu-Ang Zhou, SPIE Proc. {\bf 2559}, 47 (1995).

\bibitem{Tinkham}  M. Tinkham, {\it Introduction to Superconductivity},
(McGraw-Hill, New York, 1974), p. 250.

\bibitem{Aronson1}  I. Aronson, R. Gitterman, and B. Ya. Shapiro, Phys. Rev.
B {\bf 51}, 3092 (1995).

\bibitem{Aronson2}  I. Aronson, B. Ya. Shapiro, and V. Vinokur, Phys. Rev.
Lett. {\bf 76}, 142 (1996).

\bibitem{Martens}  J. S. Martens, {\it et al}., Appl. Phys. Lett. {\bf 58},
2543 (1991).

\bibitem{Holstein}  W. L. Holstein, L. A. Parisi, Z. Y. Shen, C. Wilker, M.
S. Brenner, and J. S. Martens, J. Supercon. {\bf 6}, 191 (1993).

\bibitem{GolosEarlyRImaging}  M. Golosovsky, and D. Davidov, Appl. Phys.
Lett. {\bf 68}, 1579 (1996).

\bibitem{GusFirstPaper}  C. P. Vlahacos, R. C. Black, S. M. Anlage, and F.
C. Wellstood, Appl. Phys. Lett. {\bf 69}, 3272 (1996).

\bibitem{XiangRxImitation}  Y. Lu, {\it et al}., Science {\bf 276}, 2004
(1997).

\bibitem{Ash+Nichols}  E. A. Ash and G. Nichols, Nature {\bf 237}, 510
(1972).

\bibitem{AnlageASC}  S. M. Anlage, {\it et al}., IEEE Trans. Appl. Supercon. 
{\bf 7}, 3686 (1997).

\bibitem{Xiang0.1micron}  C. Gao, T. Wei, F. Duewer, Y. Li, and X.-D. Xiang,
Appl. Phys. Lett. {\bf 71}, 1872 (1997).

\bibitem{Sohoo}  R. F. Soohoo, J. Appl. Phys. {\bf 33}, 1276 (1962).

\bibitem{Bryant+Gunn}  C. A. Bryant and J. B. Gunn, Rev. Sci. Instrum. {\bf %
36}, 1614 (1965).

\bibitem{Durig}  U. Durig, D. W. Pohl and F. Rohmer, J. Appl. Phys. {\bf 59}%
, 3318 (1986).

\bibitem{Gutman}  R. J. Gutman, J. M. Borrego, P. Chakrabarti and Ming-Shan
Wang, IEEE MTT-S Digest, p. 281 (1987).

\bibitem{Tabib-Azar}  M. Tabib-Azar, N. S. Shoemaker and S. Harris, Meas.
Sci. Tech. {\bf 4}, 583 (1993).

\bibitem{XiangFirstPaper}  T. Wei, X. D. Xiang, W. G. Wallace-Freedman, and
P. G. Schultz, Appl. Phys. Lett. {\bf 68}, 3506 (1996).

\bibitem{Nunes}  G. Nunes, and M. R. Freeman, Science {\bf 262}, 1029 (1993).

\bibitem{Hamers}  R. J. Hamers, and D. G. Cahill, Appl. Phys. Lett. {\bf 57}%
, 2031 (1990).

\bibitem{Seifert}  W. Seifert, E. Gerner, M. Stachel, and K. Dransfeld,
Ultramicroscopy {\bf 42-44}, 379 (1992).

\bibitem{Ichiro}  I. Takeuchi, {\it et al}., Appl. Phys. Lett. {\bf 71},
2026 (1997).

\bibitem{Stranick}  S. J. Stranick, L. A. Blumm, M. M. Kamma, and P. S.
Weiss, in {\it Photons and Local Probes}, edited by O. Marti and R. Miller
(Kluwer, Netherlands, 1995), p. 221.

\bibitem{Keilman}  F. Keilmann, D. W. van der Weide, T. Eickelkamp, R. Merz,
and D. Stakle, Opt. Commun. {\bf 129}, 15 (1996).

\bibitem{Fee+Chu}  M. Fee, S. Chu, and T. W. Hansch, Optics Commun. {\bf 69}%
, 219 (1989).

\bibitem{Dave1}  D. E. Steinhauer, C. P. Vlahacos, Sudeep Dutta, F. C.
Wellstood, and Steven M. Anlage, Appl. Phys. Letts. {\bf 71}, 1736 (1997).

\bibitem{Dave2}  D. E. Steinhauer, C. P. Vlahacos, S. K. Dutta, B. J.
Feenstra, F. C. Wellstood, and Steven M. Anlage, Appl. Phys. Lett. {\bf 72},
861 (1998).

\bibitem{GusTopo}  C. P. Vlahacos, D. E. Steinhauer, S. K. Dutta, B. J.
Feenstra, Steven M. Anlage, and F. C. Wellstood, Appl. Phys. Lett. {\bf 72},
1778 (1998).

\bibitem{ColorPics}  Color images of our microwave microscopy work are
available at http://www.csr.umd.edu/research/hifreq/micr\_microscopy.html.

\bibitem{Rebeiz}  T. P. Budka, S. D. Waclawik, G. M. Rebeiz, IEEE Trans. MTT 
{\bf 44}, 2174 (1996).

\bibitem{vdW}  D. W. van der Weide, and P. Neuzil, J. Vac. Sci. Technol. B 
{\bf 14}, 4144 (1996).

\bibitem{Ashfaq}  A. Thanawalla, {\it et al}., submitted to Appl. Phys.
Lett. (1998).

\bibitem{vdWMagnetic}  V. Agrawal, P. Neuzil, and D. van der Weide, Appl.
Phys. Lett. {\bf 71}, 2343 (1997).

\bibitem{Newman}  H. S. Newman, and J. C. Culbertson, Microwave Opt. Tech.
Lett. {\bf 6}, 725 (1993).

\bibitem{LTSEM}  R. Gerber, {\it et al.}, Appl. Phys. Lett. {\bf 66}, 1554
(1995).

\bibitem{Sudeep}  S. K. Dutta, {\it et al}., submitted to Appl. Phys. Lett.
(1998).
\end{references}
\end{document}